\documentstyle[epsf]{mn}
\newif\ifAMStwofonts    
\AMStwofontstrue

\ifoldfss
  \ifCUPmtlplainloaded \else
    \NewTextAlphabet{textbfit} {cmbxti10} {}
    \NewTextAlphabet{textbfss} {cmssbx10} {}
    \NewMathAlphabet{mathbfit} {cmbxti10} {} 
    \NewMathAlphabet{mathbfss} {cmssbx10} {} 
  \fi
  \ifAMStwofonts
    \ifCUPmtlplainloaded \else
      \NewSymbolFont{upmath} {eurm10}
      \NewSymbolFont{AMSa} {msam10}
      \NewMathSymbol{\upi}     {0}{upmath}{19}
      \NewMathSymbol{\umu}     {0}{upmath}{16}
      \NewMathSymbol{\upartial}{0}{upmath}{40}
      \NewMathSymbol{\leqslant}{3}{AMSa}{36}
      \NewMathSymbol{\geqslant}{3}{AMSa}{3E}

       \let\ge=\geqslant
    \fi
  \fi
\fi 

\ifnfssone
  \newmathalphabet{\mathit}
  \addtoversion{normal}{\mathit}{cmr}{m}{it}
  \addtoversion{bold}{\mathit}{cmr}{bx}{it}
  \newmathalphabet{\mathbfit} 
  \addtoversion{normal}{\mathbfit}{cmr}{bx}{it}
  \addtoversion{bold}{\mathbfit}{cmr}{bx}{it}
  \newmathalphabet{\mathbfss} 
  \addtoversion{normal}{\mathbfss}{cmss}{bx}{n}
  \addtoversion{bold}{\mathbfss}{cmss}{bx}{n}
  \ifAMStwofonts
    \ifCUPmtlplainloaded \else
      \UseAMStwoboldmath
      \makeatletter
      \new@mathgroup\upmath@group
      \define@mathgroup\mv@normal\upmath@group{eur}{m}{n}
      \define@mathgroup\mv@bold\upmath@group{eur}{b}{n}
      \edef\UPM{\hexnumber\upmath@group}
      \new@mathgroup\amsa@group
      \define@mathgroup\mv@normal\amsa@group{msa}{m}{n}
      \define@mathgroup\mv@bold\amsa@group{msa}{m}{n}
      \edef\AMSa{\hexnumber\amsa@group}
      \makeatother
      \mathchardef\upi="0\UPM19
      \mathchardef\umu="0\UPM16
      \mathchardef\upartial="0\UPM40
      \mathchardef\leqslant="3\AMSa36
      \mathchardef\geqslant="3\AMSa3E

       \let\ge=\geqslant
    \fi
  \fi
\fi 

\ifnfsstwo
  \DeclareMathAlphabet{\mathbfit}{OT1}{cmr}{bx}{it}
  \SetMathAlphabet\mathbfit{bold}{OT1}{cmr}{bx}{it}
  \DeclareMathAlphabet{\mathbfss}{OT1}{cmss}{bx}{n}
  \SetMathAlphabet\mathbfss{bold}{OT1}{cmss}{bx}{n}
  \ifAMStwofonts
    \ifCUPmtlplainloaded \else
      \DeclareSymbolFont{UPM}{U}{eur}{m}{n}
      \SetSymbolFont{UPM}{bold}{U}{eur}{b}{n}
      \DeclareSymbolFont{AMSa}{U}{msa}{m}{n}
      \DeclareMathSymbol{\upi}{0}{UPM}{"19}
      \DeclareMathSymbol{\umu}{0}{UPM}{"16}
      \DeclareMathSymbol{\upartial}{0}{UPM}{"40}
      \DeclareMathSymbol{\leqslant}{3}{AMSa}{"36}
      \DeclareMathSymbol{\geqslant}{3}{AMSa}{"3E}

       \let\ge=\geqslant
    \fi
  \fi
\fi 

\ifCUPmtlplainloaded \else
  \ifAMStwofonts \else 
    \def\upi{\pi}
    \def\umu{\mu}
    \def\upartial{\partial}
  \fi
\fi

\title{Long-term evolution of isolated $N$-body systems}
\author[H. Baumgardt et al.]
       {Holger Baumgardt,$^1$ Piet Hut,$^2$ Douglas C.~Heggie,$^3$\\
        $^1$Department of Astronomy, School of Science, The University of Tokyo,
            7-3-1 Hongo, Bunkyo-ku, Tokyo 113-0033, Japan\\
        $^2$Institute for Advanced Study, Princeton, NJ 08540, USA\\
        $^3$Department of Mathematics and Statistics, University of Edinburgh, 
         King's Buildings, Edinburgh EH9 3JZ, UK}
\date{Accepted .
      Received ;
      in original form }

\pagerange{\pageref{firstpage}--\pageref{lastpage}}
\pubyear{2000}

\begin{document}

\maketitle

\label{firstpage}

\begin{abstract}
We report results of $N$-body simulations of isolated star clusters, performed up to the point where the
clusters are nearly completely dissolved. Our main focus is on the post-collapse evolution of these 
clusters. We find that after core collapse, isolated clusters 
evolve along nearly a single sequence of models whose properties are independent of the initial 
density profile and particle number.   
Due to the slower expansion of \mbox{high-$N$} clusters, relaxation times become almost
independent of the particle number after several core collapse times, at least for the particle
range of our study. As a result, the dissolution
times of isolated clusters exhibit a surprisingly weak dependence on $N$.

We find that most stars escape due to encounters between single stars inside the half-mass radius of the
cluster. Encounters with binaries take place mostly in the cluster core and account for roughly
15\% of all escapers. 
Encounters between single stars at intermediate radii are also responsible for
the build up of a radial anisotropic velocity distribution in the halo. 
For clusters undergoing core oscillations, escape due to binary stars
is efficient only when the cluster center is in a contracted phase. 
Our simulations show that it takes about $10^5$ $N$-body time units until the global anisotropy 
reaches its maximum
value. The anisotropy
increases with particle number and it seems conceivable that isolated star clusters become vulnerable to
radial orbit instabilities for large enough $N$. However, no indication for the onset of such instabilities was 
seen in our runs.  
\end{abstract}


\section{Introduction}

It is one of the great challenges of numerical astrophysics to understand the dynamical evolution of globular
clusters and other collisional systems (as e.g.\ galactic nuclei;  galaxy clusters). So far however,
it is not possible to perform simulations of such rich systems by direct summation techniques
and one has to rely on a statistical modeling. This is still the case, despite recent progress in the development
of faster computers, like e.g.\ the GRAPE computer series (Makino 2001, 2002). 

Statistical modeling of globular clusters includes different techniques like Fokker-Planck, Monte Carlo and Gaseous 
model simulations. There has been substantial progress in recent years in improving the reality of such models and 
comparing them with $N$-body models or comparisons between different statistical methods show good agreement 
in many cases (Takahashi \& Portegies-Zwart 2000, Giersz \& Spurzem 2000). 
Nevertheless the validity of such codes for real globular clusters should be checked further, due to
the complexity of the physical processes involved and 
the fact that the gap between the largest published $N$-body models (which are of order $3 \cdot 10^4$), and 
the number of stars in real globular clusters (roughly $N = 10^6$) is still large. 

Isolated star clusters are an ideal environment to test star cluster evolution. These systems are 
sufficiently simple,
but nevertheless show many phenomena that are also seen in real globular clusters. 
Although simulations of isolated clusters are
not directly applicable to real stellar systems, they are nevertheless
highly relevant.  Borrowing an example from stellar evolution,
such calculations play the role that polytropic models have
played there historically.  While rather unrealistic physically, they
did shed considerable light on the basic aspect of stellar structure,
and in addition they served as useful comparison material when trying to
interpret the results of the far more detailed numerical evolution models.
Models for isolated clusters have the same dual role in stellar dynamics, they 
show some of the major physical effects occurring in long-term
dynamical evolution, and provide templates against which more detailed
models can be interpreted. A large amount of research has
therefore been devoted to the study of isolated clusters. 

It has been known for a long time that star clusters undergo a contraction of their center as a result of heat 
transfer from the cluster core to the halo (Antonov 1962, Lynden-Bell \& Wood 1967). For isolated, single-mass
clusters, this core collapse takes roughly 17 initial relaxation times until completion (Takahashi 1995),
and the core collapse evolution is thought not to change very much in the presence of an
external tidal field (Giersz \& Heggie 1997). Lynden-Bell \& Eggleton (1980) found that the core collapse
proceeds from the outer to the inner parts and leaves behind a power-law debris of material 
with density exponent in 
the range $\alpha = -2$ to -2.5. Although core collapse strongly influences the structure of the inner parts,
one usually assumes that the overall properties of the system are determined more by the half-mass
radius, and the core adjusts itself in post-collapse so as to balance the tendency of the halo to recollapse.

Core collapse will be reversed if a central energy source is present (H\'enon 1975), and it is
normally assumed that binary stars, which are either primordial or form during core collapse 
and harden as a result of encounters
with field stars, provide this heat source (Aarseth 1971, Heggie 1975) in real star clusters. 
The number of binaries necessary to drive the cluster evolution is found to be quite small.
Goodman (1984) estimated from his gaseous model calculations that their number drops
with the number of cluster stars as $N_b \propto N^{-0.3}$ and that in the mean only $N_b = 0.5$ 
binaries are necessary
to drive the evolution of a cluster with $N = 10^6$ stars. This has been partly confirmed by
Giersz \& Heggie (1994b) in their $N$-body simulations of small-$N$ clusters.
They found that only few binaries are present in the core, but their number seemed to increase with
the particle number. They noted however that some of their binaries might be temporary ones which 
contribute nothing to the energy generation of the core. 

Stars scattered out of the cluster center create an radially anisotropic
velocity profile in the cluster halo (Larson 1970). Cohn (1984) found that a cluster 
starts to become anisotropic in its outer parts already early on in the pre-collapse phase, while
Spitzer \& Shapiro (1972) have argued that the anisotropy extends down into the
core as a consequence of its collapse.
This was later confirmed by Takahashi (1995) and Drukier et al.\ (1999) for the late stages of
core collapse. For post-collapse clusters, Takahashi (1996) found that they are
isotropic in their center and the anisotropy increases monotonically
towards the halo.

Encounters of stars in the cluster center also lead to the escape of stars. H\'enon (1960) has shown 
that stars in isolated systems do not escape by distant encounters, but only due to close encounters with
other stars.
Despite their small numbers, it is generally believed that binaries play an important role in the
production of escapers and contribute to the development of an anisotropic velocity profile in 
the cluster halo.

So far, most simulations of isolated star clusters were concerned only with the early stages
up to core collapse or the immediate re-expansion phase. Nearly nothing is known about how clusters 
evolve when substantial mass-loss sets in. One reason for this neglect is that the escape of 
stars from isolated clusters happens very slowly and on a timescale which increases rapidly
as the clusters expand. In addition, large-angle scatterings are neglected in Fokker-Planck 
methods and can only be studied by $N$-body or Monte Carlo simulations.

Hence, although highly desirable, it is difficult to verify the predictions of Fokker-Planck or 
Gaseous models by direct $N$-body simulations.
We therefore present in this paper the first $N$-body simulations of medium sized isolated star clusters which
follow the evolution until near complete disintegration.

\section{Description of the runs}

We simulated the evolution of clusters containing between $N = 128$ and 8192 equal-mass stars,
increasing the particle numbers by successive factors of 2. Our clusters followed
Plummer profiles initially and we did not impose a cut-off radius when the Plummer 
models were created. 
Small-$N$ runs were made with the collisional Aarseth $N$-body code NBODY6
(Makino \& Aarseth 1992, Aarseth 1999) on a Pentium PC workstation. The 8K run 
was made with the NBODY6++ code (Spurzem 1999, Spurzem \& Baumgardt 2002),
which is a parallelized version of NBODY6. It was made on a Cray T3E parallel
computer using 8 processors. Though these $N$ are not large by the standards of
current more realistic simulations, the computations are at least as
time-consuming.

During the runs, we recorded the positions and velocities of all stars at regular intervals. In the
first 100 $N$-body time units, the output time $\delta T$ was equal to one initial crossing time.
Later, time-intervals were equally spaced in $\log  T$. At each 
data output, we determined the number of bound stars, the lagrangian radii of the members
(measured relative to the density center) and the velocity structure of the bound stars. 
We also calculated a scaled time according to 
\begin{equation}
T_{Scale} = \int_0^{T_{NBODY}} \left( \frac{r_h(0)}{r_h(t)} \right)^{3/2} dt \;\; ,
\end{equation} 
where $r_h$ is the half-mass radius of the bound cluster. This time takes the expansion
of the cluster into account (but not the mass-loss), and measures how much time has passed in comoving
coordinates. Data was also stored every 2.828 timeunits in scaled time. 

The runs were performed for a maximum time of $T_{Max} = 10^{16}$ $N$-body units.
In order to follow the evolution up to the point where nearly all stars are unbound, we had to
rescale the whole system when its size became too large. Otherwise the simulations would have come to a
halt due to numerical difficulties in NBODY6.
Rescaling was done by calculating the sum of the potential energies of
all stars, excluding the internal energy of regularised binaries. The positions and velocities of all
stars were then rescaled such that the total potential energy after rescaling was equal to
$E_{Pot} = -1/2$. The cluster centers were also moved back to
the origin and the mean cluster velocity was subtracted from all stars. This recentering did not
influence the cluster membership. In general, the whole integration could be covered 
by two rescalings.

Bound stars were determined in the following way: We first calculated the potential energy of
each star with respect to all other stars. Kinetic energies were then calculated relative to the
motion of the cluster center, which was taken from a previous iteration, or, at the start of the run,  
assumed to be zero. The mean cluster motion was then determined from the bound stars
(i.e.\ those with total energies $E < 0$) and in the following iteration potential energies were calculated
with respect to the bound stars only. This method was repeated until a stable solution for the
cluster motion and the number of bound stars was found. From the cluster members, we 
determined the position of the density center and the lagrangian radii, using the method of 
Casertano \& Hut (1985). Unbound stars were not removed from the simulations, although for 
the analysis in this paper we will restrict ourselves to the bound stars. We note that throughout the paper, 
the term cluster will be used for the system of bound stars only. We also note that using unbound stars
for the calculation of the potential energy or the position of the density center would make little 
difference to our results, since 
they contribute little to the potential energy and also have only a small
weight in the calculation of the density center.
 
\begin{table}
\caption[]{Details of the performed $N$-body runs.}
\begin{tabular}{rcl@{$\,\cdot$}rcrl@{$\,\cdot\,$}r}
\noalign{\smallskip}
\multicolumn{1}{c}{$N$}& \multicolumn{1}{c}{$N_{Sim}$} &  
\multicolumn{2}{c}{$T_{Finh}$} & \multicolumn{1}{c}{$<\!N_{Tmax}\!>$} & \multicolumn{1}{c}{$T_{CC}$} & \multicolumn{2}{c}{$T_{Half}$} \\
\noalign{\smallskip}
  128 &  64 &  2.8 & $10^{12}$ & $-$  &  70.7 &  6.75 & $10^{4}$\\
  256 &  64 &  1.3 & $10^{14}$ & $-$  & 104.7 &  6.70 & $10^{4}$\\
  512 &  40 &  1.2 & $10^{15}$ & 12.7 & 169.7 &  7.56 & $10^{4}$\\
 1024 &  20 &  1.0 & $10^{16}$ & 28.5 & 360.3 &  9.05 & $10^{4}$\\
 2048 &  10 &  1.0 & $10^{16}$ & 49.4 & 587.5 &  1.39 & $10^{5}$\\
 4096 &  5  &  1.0 & $10^{16}$ &100.5 &1031.2 &  2.29 & $10^{5}$\\
 8192 &  1  &  1.0 & $10^{16}$ &182.0 &1839.2 &  3.79 & $10^{5}$\\
\end{tabular}
\end{table}

Table 1 gives an overview of the simulations. It shows the number of cluster
stars $N$, the number $N_{Sim}$ of simulations performed, 
the time $T_{Finh}$ when half the simulations had stopped (explained below), the number of stars still
bound at $10^{16}$ \mbox{$N$-body} time units and the core collapse and half-mass times of the models.  

Runs were terminating before $T_{Max}$ due to either difficulties in the 
numerical integration or because our routine for the membership determination couldn't find bound stars. 
This happened for example when a cluster consisted of only two or three stars and these were merged
by NBODY6 to form an hierarchical system. Runs which finish before $T_{Max}$ might bias our results,
so results obtained for $T \ge T_{Finh}$ should be treated with caution. However, Table 1 shows that
only results for small-$N$ clusters might be biased and these also only for the very late stages of
the evolution. 

Throughout
the paper we will use $N$-body units, so the constant of gravitation, initial cluster mass and energy
are given by $G=1$, $M=1$ and $E_C=-0.25$ respectively.

\section{Results}

\subsection{Pre-collapse evolution}

We will start our analysis by discussing the pre-collapse evolution of the clusters. Fig.~1
compares the evolution of different models after the $N$-body time units were divided by the 
half-mass relaxation time (Spitzer 1987)
\begin{equation}
T_{RH} = 0.138 \; \frac{\sqrt{N} \; \; r_h^{3/2}}{\sqrt{m} \; \sqrt{G} \; ln \gamma N}  \;\; ,
\end{equation}                                                 

where $N$ is the initial number of cluster stars, $m$ their mean mass, $r_h$ the initial 
half-mass radius of the 
cluster, $\gamma$ a constant in the Coulomb logarithm (discussed below), and $G$ the constant of gravitation.

The bottom panel shows lagrangian radii containing between 3\% to 70\% of the bound stars,
calculated by averaging the radii from individual runs. The collapse of the clusters and the 
following re-expansion can be clearly seen. Core collapse is achieved in approximately 17 initial relaxation 
times, which agrees well with published results from anisotropic Fokker-Planck runs (Takahashi 1995, 
Drukier et al.\ 1999).  For most of the
pre-collapse phase good agreement in the evolution of the lagrangian
radii for different $N$ can be seen. Significant deviations occur only shortly before core collapse,
when heat input by binaries starts to influence the core evolution. Giersz \& Spurzem (1994) found
that high-$N$ clusters undergo a core collapse which is deeper and happens at later times since 
binaries in low-$N$ clusters have to create less energy to stabilise the cluster core from 
further contraction. Goodman (1987) found in his gaseous model calculations that high-$N$ clusters have 
a smaller core radius also in the
post-collapse phase. As can be seen, these results agree very well with the results of our $N$-body 
calculations. Fig.\ 1 shows 
that the core radii of high-$N$ models remain smaller during the post-collapse phase, while
their intermediate radii, probably as a result of the stronger energy generation in the core, 
are larger.
\begin{figure}
\epsfxsize=8.3cm
\begin{center}
\epsffile{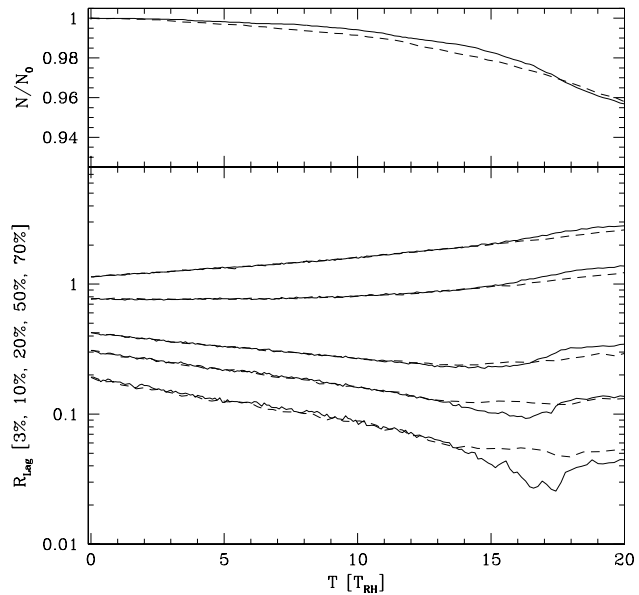}
\end{center}
\caption{Pre-collapse evolution of clusters containing $N=1024$ (dashed) and $N=4096$ (solid line) stars.
  The bottom
 panel shows the evolution of the lagrangian radii, the top panel shows the evolution of
  the bound mass. Time is scaled by the initial relaxation time. The lagrangian radii
   evolve in the same way in both models until shortly before core collapse.}
\label{pre_cc}
\end{figure}

The outer radii are expanding right from the 
start of the simulations. In agreement with Cohn (1984), we find that the halos 
become anisotropic already during the pre-collapse phase (Fig.\ \ref{an1}), so the pre-collapse
expansion is at least partially
driven by stars ejected from the cluster cores. These stars must be ejected by two-body encounters
between single stars, as no binaries are present in the clusters before core collapse. 

Comparing the evolution of models with different $N$ allows us
to determine the value of the Coulomb logarithm used in eq.\ 2. From a fit of the lagrangian
radii, we typically obtained values between $\gamma = 0.05$ and 0.2, depending on which
radii and particle numbers are compared. These values are similar to the value of $\gamma = 0.11$
obtained by Giersz \& Heggie (1994a) from a comparison of $N = 500$ and $N=2000$ models.
Since the number of our simulations is relatively small, these differences are probably within the
statistical error.

The upper panel of Fig.\ \ref{pre_cc} depicts the mass-loss. The mass-loss rate
increases monotonically towards core collapse and about 3\% of the mass is lost in pre-collapse.
There is a trend that the 
4K-model loses mass at a slightly smaller rate in the beginning, and this effect can also be seen if we 
compare other models.
The reason could be that stars in isolated models escape only by very close encounters,
for which H\'enon (1960) found that they lead to a scaling of the lifetime proportional to $N$ 
rather than $N/\log(N)$.

\subsection{Post collapse expansion}

The post-collapse expansion of an isolated cluster is usually assumed to be self-similar.
In this case, its evolution is characterized by the following two equations:
\begin{eqnarray}
\frac{d M(t)}{d t} & = & - k_1 \; \frac{M(t)}{t_{rh}} \\ 
\nonumber \frac{d r_h(t)}{d t} & = & k_2 \; \frac{r_h(t)}{t_{rh}}
\end{eqnarray}

If the variation in the Coulomb logarithm of the relaxation time can be neglected, the solution for
the mass-loss and the half-mass radius is given by:
\begin{eqnarray}
M(t) & = & M_0 \; (t/t_0)^{-\nu} \\
\nonumber  r_h(t) & =  & r_{h0} \; (t/t_0)^{(2+\nu)/3} \;\; ;
\end{eqnarray}
where $t_0$ is the time of core collapse and $\nu$ a constant. Goodman (1984) obtained values of $\nu$
ranging from 0.0258 to 0.100 for the evolution of his homologous sphere. Similar values were also
found by Giersz \& Heggie (1994b) in their $N$-body simulations of small-$N$ clusters.
On the other hand, Drukier et al.\ (1999) found strong deviations from a $r \propto t^{2/3}$ scaling 
for clusters with $N \ge 8000$ stars
in their anisotropic Fokker-Planck simulations, especially for the outer radii.  
They attributed these changes to the buildup of a radially anisotropic velocity distribution
in the cluster halos.

Fig.\ \ref{ged} shows the evolution of the lagrangian radii as a function of $N$-body time in our
simulations.
We have shifted the $N$-body time units such that all clusters go into core collapse at the same time
as the 4K model.
Due to their smaller relaxation times, low-$N$ clusters need less time to reach core collapse
and also expand quicker after core collapse, which can be seen most clearly in the
evolution of the outer lagrangian radii.
\begin{figure}
\epsfxsize=8.3cm
\begin{center}
\epsffile{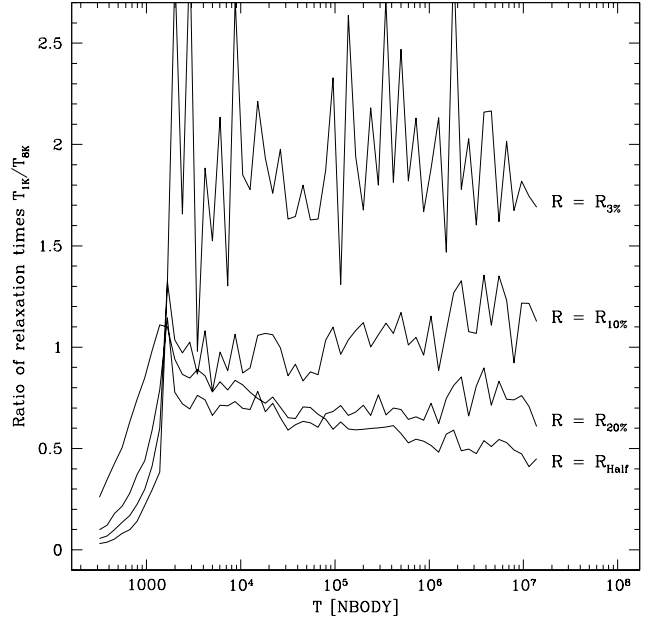}
\end{center}
\caption{Ratio of the relaxation times of the 1K and 8K models at different radii. The 8K model
 has a smaller relaxation time in the core but a larger relaxation time in the halo. Relaxation times
 are approximately the same at the 10\% radius, showing that the cluster expansion is driven from
  around this radius.}
\label{rrel}
\end{figure}

A strong increase of the outer radii might be expected since the halo density in a Plummer model drops off like
$\rho \propto r^{-5}$, while for an isolated cluster with mass-loss one expects
to find an equilibrium relation $\rho \propto r^{-3.5}$ (Spitzer \& Shapiro 1972). The build up of this halo
will cause the outer radii to expand and might also explain 
the increase seen by Drukier et al. (1999, their Fig. 6). 

The quick expansion of low-$N$ clusters
cannot be sustained indefinitely however, since, as a cluster expands, its relaxation
time increases with it. Hence, the expansion of a low-$N$ cluster cannot go beyond a point where 
its relaxation time reaches the relaxation time of an high-$N$ cluster. After some time has 
passed in post-collapse expansion, one would therefore 
expect that high-$N$ clusters have smaller radii which compensate their large particle
numbers, and that all clusters expand with the same rate. 
It can be seen in Fig.\ \ref{ged} that this is nearly the case.

Interestingly, from a comparison of different models, one can find a
Lagrangian radius at which the relaxation time (estimated in a manner
to be described) is the same for all models at the same time.  In a
sense this can be interpreted as the radius which drives the
expansion.  In order to estimate the relaxation time at a given
Lagrangian radius $r$ we use eq.(2) with $r_h$ replaced by $r$.  This does
not take into account the fact that the mass within radius $r$ varies
with $r$, but since our purpose is to compare different models at the
Lagrangian radius corresponding to the same fractional mass, this is
not important.  
Fig.\ \ref{rrel} shows the ratio of the relaxation times of the 1K and 8K clusters 
at 4 different radii as a function of time. Throughout the post-collapse evolution, both clusters 
have similar relaxation times at the 10\% radius, although there is some tendency towards 
larger radii towards the end. Similar results are obtained from a comparison
of other $N$. The expansion
of isolated clusters may therefore be thought of as being driven from around the 10\% radius. The 
cluster cores will
adjust themselves and their energy generation to deliver the necessary energy, while relaxation
times at radii outward from the 10\% radius are too large to influence the expansion much. 
\begin{figure}
\epsfxsize=8.3cm
\begin{center}
\epsffile{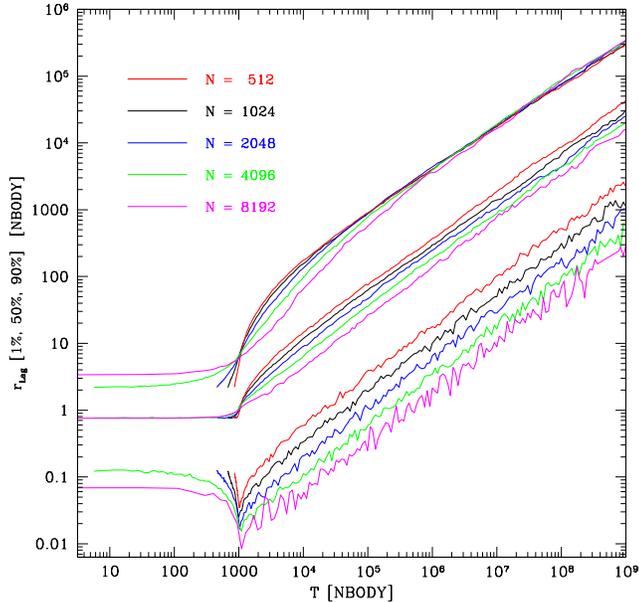}
\end{center}
\caption{Evolution of the lagrangian radii for clusters with different initial particle numbers.
 Times are shifted such that all clusters go into core collapse at the same time. Low-$N$
 clusters expand quicker after core collapse, and due to this faster expansion
  their relaxation times reach those of high-$N$ clusters.
    After $10^5$ $N$-body time units, all clusters do therefore expand with the same rate.}
\label{ged}
\end{figure}

\begin{figure}
\epsfxsize=8.3cm
\begin{center}
\epsffile{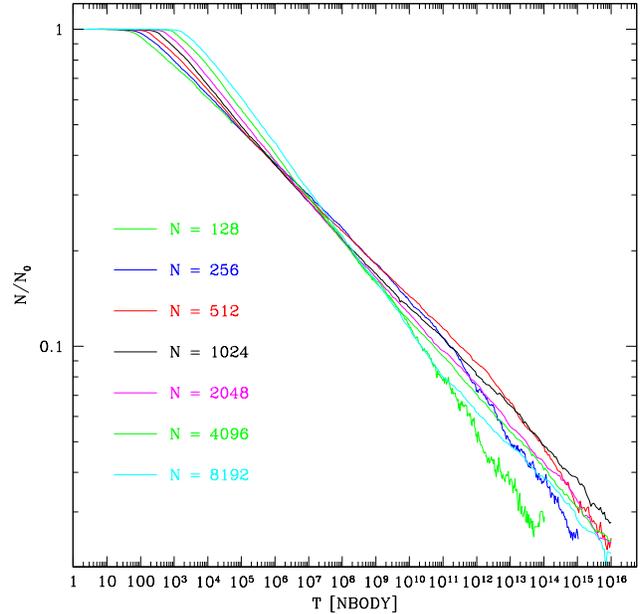}
\end{center}
\caption{Fraction of bound mass as a function of $N$-body time. Shown are clusters
  containing between $N = 128$ to $N = 8192$ stars. Although low-$N$ clusters have initially
  smaller relaxation times and start losing mass earlier, curves for different $N$ meet each other
  at about $10^7$ $N$-body time units. Afterwards, high-$N$ models contain a smaller amount
  of mass at any given time, except for models with $N<512$.}
\label{mloss}
\end{figure}

\begin{figure}
\epsfxsize=8.3cm
\begin{center}
\epsffile{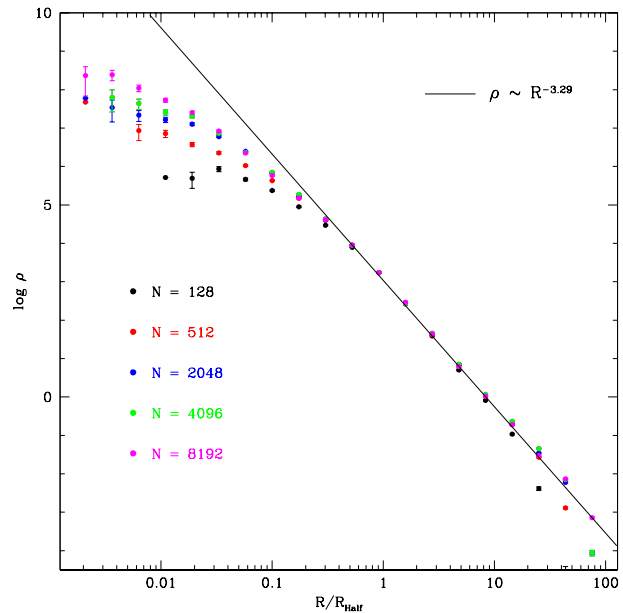}
\end{center}
\caption{Density profiles for clusters starting with $N = 128$
   to $N = 8192$ stars. Radii are given in units of the half-mass radius and densities
  are adjusted to be the same for all $N$ at the half-mass radius. The solid line shows
   a density profile proportional to $\rho \propto R^{-3.29}$. It gives a good fit to the
   halo profiles, especially for large $N$.}
\label{dp1}
\end{figure}

Although it is clear from Fig.\ \ref{ged} that the post-collapse expansion cannot be strictly
self-similar, the inner lagrangian radii up to the half-mass radius expand regular
enough to be fitted by a relation like eq.\ 3.
Fitting the evolution of different lagrangian radii and the mass-loss
curve between $10^3$ $N$-body time units and $10^8$ $N$-body time units 
($10^4$ to $10^8$ for $N \ge 4096$) 
gives the $\nu$-values shown in Table~2.  
For all $N$, $\nu$ drops with increasing distance from the cluster 
center, so that the core radii show the fastest expansion. The reason
for this will be discussed in the next section. 

Our values for the expansion of the half-mass radius agree fairly well with the values found 
by Goodman (1984) and Giersz \& Heggie (1994b), despite the fact that different times were
used by Giersz \& Heggie (1994b) to obtain them and the scatter in $\nu$ is fairly large
for our high-$N$ runs. Unfortunately no comparison values are available for the inner
lagrangian radii.
The best agreement with the $\nu$-values derived from the mass-loss curve is achieved for the
10\% and 20\% lagrangian radii. Since escapers are created mainly
between these radii (see section 3.4), a close correspondence in the behavior with time 
might be expected.

Another interesting result of Table 2 is that the $\nu$ values derived from the mass-loss
curve increase with $N$, which means that high-$N$ clusters lose mass faster
than clusters with smaller particle numbers. This is in contrast to a naive
application of standard relaxation theory, in which it is assumed that $\nu$ is 
independent of $N$ (eqs.~4). A large part of this discrepancy can be explained by
the higher concentration of high-$N$ clusters, which breaks the homology and makes
$k_1$ (eq.~3), and therefore $\nu$, increase with $N$.

Plotting the bound mass as a function of time (Fig.\ \ref{mloss})
shows that low-$N$ clusters start losing
mass quicker due to their earlier core collapse, but that the increase in $\nu$ is 
sufficient to reverse this after about $10^7$ $N$-body time units.
Afterwards, high-$N$ runs contain a smaller fraction of their initial mass at a
given $N$-body time, except for runs with $N<512$ which are affected by the
increasing incompleteness towards the end. 

Finally, Fig.\ \ref{ged} (and also Fig.\ \ref{esc5}) shows that our 8k-model undergoes core oscillations up to 
about $10^4$ $N$-body time units, when the number of cluster stars drops below 6700. The minimum
required mass for core oscillations seems therefore to be of order 7.000, which agrees
very well with results obtained by Goodman (1987) from his conducting gas sphere calculations.
Only slightly larger values were obtained by Drukier et al.\ (1999) and Breeden et al.\ (1994), who
found a lower limit of $N \ge 8000$.

\subsection{Density profile in post-collapse}

\begin{table}
\caption[]{The parameter $\nu$ determined from the mass-loss and the expansion rate of 
four different lagrangian radii as a function of the particle number $N$. Closest agreement
with a self-similar expansion is achieved if the expansion of the 10\% radius is compared
with the mass-loss rate.} 
\begin{tabular}{crrrrrrr}
\noalign{\smallskip}
  N     &   128 &  256  & 512  & 1024 & 2048 & 4096 & 8192 \\[+0.1cm]
$\nu_{M}$& 0.10 &  0.11 & 0.11 & 0.13 & 0.13 & 0.14 &  0.14 \\[+0.1cm] 
$\nu_{r3}$ & 0.16 & 0.16 & 0.17 & 0.16 & 0.15 & 0.17 &  0.22 \\    
$\nu_{r10}$& 0.13 & 0.11 & 0.14 & 0.14 & 0.13 & 0.11 &  0.13 \\
$\nu_{r20}$& 0.08 & 0.12 & 0.13 & 0.11 & 0.08 & 0.05 &  0.09 \\
$\nu_{r50}$& 0.08 & 0.09 & 0.07 & 0.02 & 0.01 & 0.06 &  0.09 \\   
\end{tabular}
\end{table}

\begin{figure}
\epsfxsize=8.3cm
\begin{center}
\epsffile{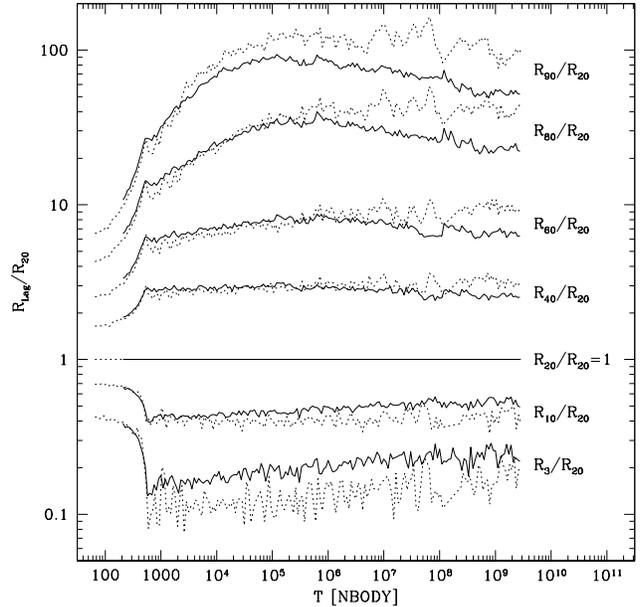}
\end{center}
\caption{Ratio of the lagrangian radii relative to the 20\% radius for clusters with $N=8192$
 (dashed line) and $N=2048$ (solid line) stars. The time in the 8K model is rescaled by the
  relaxation time to that of the 2K model.
  Halos expand after core collapse until their build-up is reversed by the removal of
   weakly bound stars from the recoil motion of the center due to escaping stars.
   The cores are expanding slowly throughout the simulation because of increased binary activity
  in the center.}   
\label{sesi}
\end{figure}

Fig.\ \ref{sesi} depicts 
the relative sizes of the different radii for the 2K and 8K models in greater detail.
The time of the 8K model is rescaled by the initial relaxation time to match the relaxation
time of a 2K model, and radii are scaled by the actual 20\% radius. 
The build up of the halo is happening very slowly for both $N$ and  
is complete only after about $10^5$ $N$-body time units, corresponding to $10^3$ initial relaxation times. By this 
time, the structure of
the clusters outside the 20\% radius is very similar for both $N$. Afterwards, the outermost radii in 
the 2K model start to decrease. This decrease is caused by the 
recoil motion of the clusters due to escaping stars. Due to this recoil, weakly bound
stars in the cluster halos gain positive energies relative to the cluster centers and escape.
The recoil motion is larger and more erratic for low-$N$ clusters 
since a single star carries a larger fraction of the total cluster mass. Their halos will therefore
be truncated at smaller radii. 

In contrast to the halo, the post-collapse profile inside the half-mass radius is established already 
during core collapse and changes only little afterwards. The core radii of high-$N$ clusters 
are more concentrated relative to the 20\% radius than those of low-$N$ clusters. 
Giersz \& Spurzem (1994) found that in low-$N$ models binaries have a higher chance to form 
and have to release less energy to prevent the core from contracting. They studied mainly the 
pre-collapse phase, however, given the similar density profile, the same conditions should prevail 
in the post-collapse phase. The larger core size of low-$N$ clusters is therefore due to the
stronger binary activity in their centers. Consequently, as the clusters evolve and lose mass, their
core radii increase with time. 

Fig.\ \ref{dp1} shows the density profiles at the time the cluster halos reach their maximum expansion
relative to the 20\% radius. Radii are given in units of the half-mass radius and densities
are adjusted such that they are the same at the half-mass radius.
In the halo, the density profiles of all clusters follow power-law distributions $\rho \propto R^{-\alpha}$
with slope $\alpha = 3.29$. Deviations from this occur only in the outermost bins,
where small-$N$ clusters have steeper profiles because of the recoil motion from escaping stars,
which effectively truncate the clusters at a certain radius.

Our value for $\alpha$ is not far from the theoretical prediction by Spitzer \& Shapiro (1972), who
found $\alpha = 3.5$ for the halo density of an isolated star cluster. Similar slopes
were also found by Lightman \& Shapiro (1978) and by Takahashi (1996) in his Fokker-Planck
simulations of anisotropic post-collapse clusters. 

The transition between the halo and the inner parts happens at around $R = 0.4$
half-mass radii. Inward from there, clusters have shallow density profiles which become steeper
for increasing $N$. Inagaki \& Lynden-Bell (1983) and Takahashi (1995) found that the
inner profiles can be fitted by power-law distributions with slopes of 
$\alpha = 2.0$ and $\alpha = 2.23$ respectively. Our profiles do not strictly follow
power-laws, but can be fitted by them over a restricted range in $R$. For the 4K and 8K models, 
an increase of $\rho \sim R^{-2.33}$ gives a good fit between 0.02 and
0.3 half-mass radii, which is not too different from the result of Takahashi (1995).
Given the relatively small number of stars in our simulations, and the fact that
the profiles have not reached a limiting profile even for our largest $N$,
it might be possible that a single power-law profile is reached in the centers 
of large-$N$ clusters. 

\subsection{Evolution along a single sequence of models ?}

In the expansion phase after core collapse, the space that the clusters initially occupied
becomes a vanishing fraction of their actual volume. One might therefore ask if the density 
profile after a large enough time has passed still depends on the initial profile.  
In order to test this assumption, we made a set of 8 comparison runs which started from King $W_0 = 3.0$ 
profiles and had $N=1024$ stars initially. 

\begin{figure}
\epsfxsize=8.3cm
\begin{center}
\epsffile{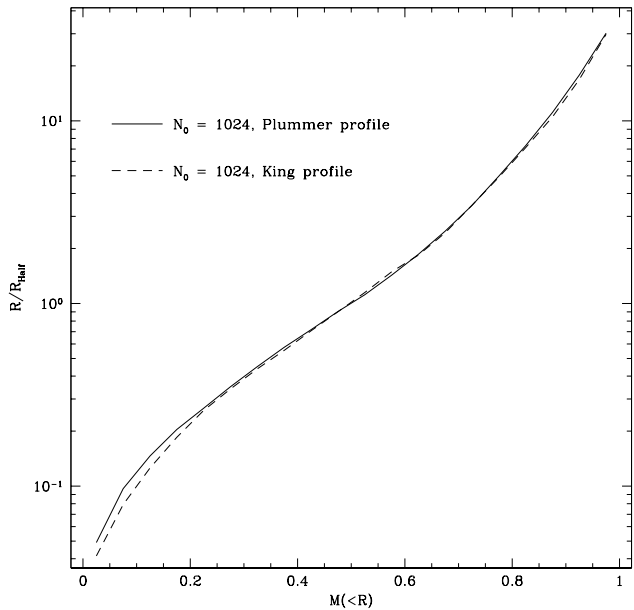}
\end{center}
\caption{Cumulative mass distribution at half-mass time for clusters starting with different
 density profiles. 
 Clusters starting with Plummer profiles are shown by a solid line, clusters starting with
 King $W_0 = 3.0$ profiles by a dashed line. The distributions at half-mass time agree 
 approximately with each other
 since the cluster expansion has erased the memory of the starting condition.}  
\label{dp2}
\end{figure}

Fig.\ \ref{dp2} compares the cumulative mass distribution of the King-models with the $N = 1024$ 
Plummer-model, after both have lost half their initial mass (note that the clusters have expanded 
by about a factor of 100 by this time). 
A cumulative distribution has the advantage that it better shows the behavior of the bulk of the stars.
It can be seen that the mass profiles of both clusters agree very well with each other.
Similar comparisons at other times give the same agreement, provided the time is much larger than the 
core collapse time. 
The density distribution of isolated clusters long after core collapse becomes therefore independent of 
their initial density profile.

\begin{figure}
\epsfxsize=8.3cm
\begin{center}
\epsffile{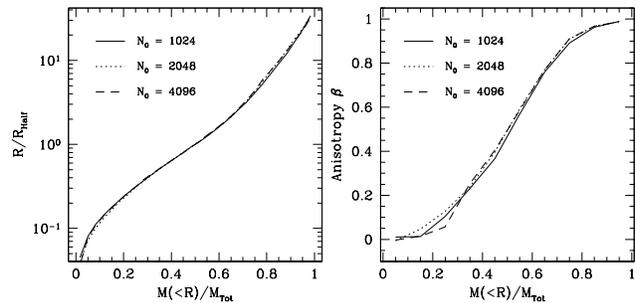}
\end{center}
\caption{Distribution of density and anisotropy $\beta$ of clusters starting with different initial 
 particle numbers by the time 512 stars are bound. Radii are scaled by the half-mass
  radius. $\beta$ is calculated as explained in section 3.7.
    Both the mass and anisotropy profiles are independent of the initial 
      particle number $N_0$.}
\label{dp3}
\end{figure}

Similarly one might ask whether the density profile or any other cluster parameter at a given particle 
number still depends on the
initial number of cluster stars. Fig.\ \ref{dp3} depicts the density and anisotropy profile as 
a function of
enclosed mass by the time 512 stars remain bound, for clusters starting between $N_0 = 1024$ 
to $N_0 = 4096$ stars. The density and anisotropy profiles agree well with each other for any 
radius, and similar plots at other times also give excellent 
agreement between the runs. 
\begin{figure}
\epsfxsize=8.3cm
\begin{center}
\epsffile{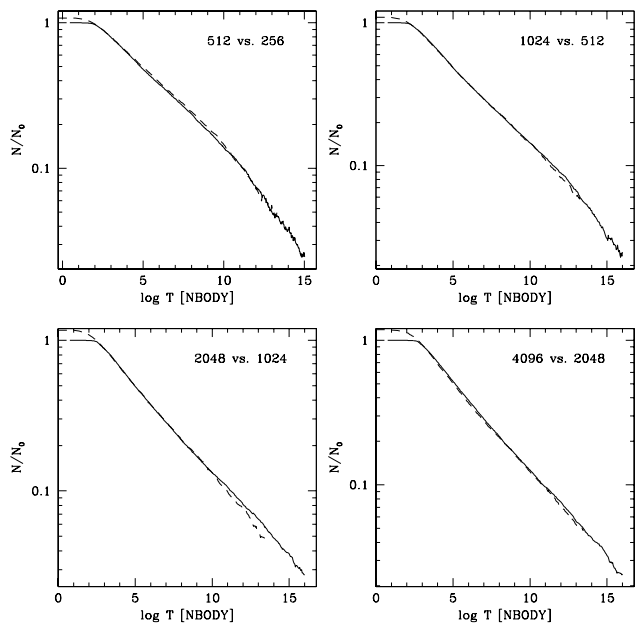}
\end{center}
\caption{Evolution of the number of bound stars for different clusters.
 Radii are rescaled such that the size of high-$N$ clusters when they have 45\% bound stars
   is equal to the size of low-$N$ clusters with
    90\% bound stars. $N$-body time units are scaled by $r_h^{3/2}$. After this rescaling, the
     bound mass evolves similar in both cases, showing that the evolution is independent of the
      initial particle number.}
\label{sti1}
\end{figure}

We conclude that the cluster structure becomes
independent of the initial particle number and density distribution after enough time 
has passed and is solely a function of the actual particle number. 
Clusters in late post-collapse can therefore be characterised by only two parameters,
as e.g. the current particle number and the half-mass radius. 
This would imply that isolated clusters evolve along a single sequence of profiles
and that the evolution of clusters which started from different
initial configurations is similar.

To check this assumption, we have a look at the evolution of bound mass with time. 
If clusters
evolve along a single sequence, the evolution of low-$N$ clusters should be similar to that
of high-$N$ clusters when they have reached the same number of stars. Fig.\ \ref{sti1}   
compares the mass-loss rate of small-$N$ clusters with clusters that had initially
twice as many stars. High-$N$ runs are rescaled such that the cluster sizes when 
45\% of the stars remain bound are equal to the sizes of the low-$N$ clusters when 90\% of their 
stars are bound. The time is also shifted such that clusters reach 90\% and 45\% bound 
stars at the same time. It can be seen that we obtain a very good fit for the evolution of bound
mass with time. The fit is best when 
many simulations were made as e.g. in cases with $N=512$ and $N=1024$. Deviations
occur, if at all, only at the end of the runs when the particle numbers are small.
We conclude that isolated clusters evolve along a single sequence of models.

\subsection{Escape of stars}

It was shown by H\'enon (1960) that stars can escape from isolated clusters only due to close
encounters with other stars or binaries. If close encounters are the dominant escape mechanism,
most stars should escape from the inner cluster regions,
where the density of stars is highest and close encounters between stars should happen most often,
though the potential well is deepest.
\begin{figure}
\epsfxsize=8.3cm
\begin{center}
\epsffile{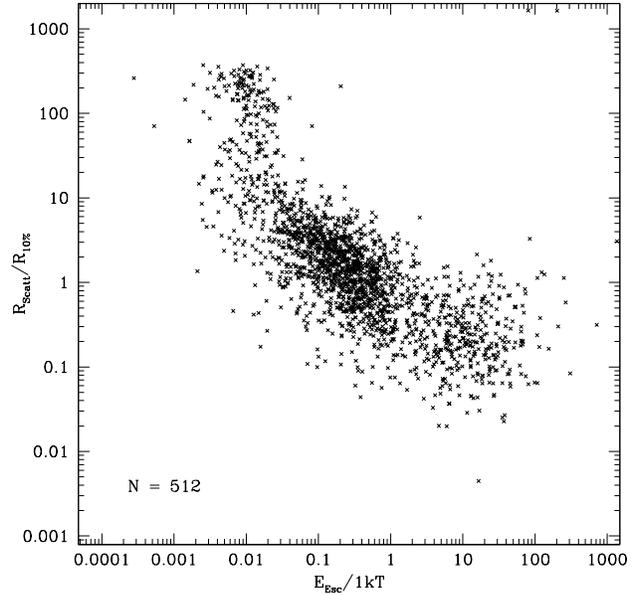}
\end{center}
\caption{Distribution of escape energies against radius where stars become unbound for the $N =
   512$ star clusters. Plotted are all stars that escape before the half-mass time. The escape energies
  are measured after the stars have reached twice the
   scattering distance. Most stars escape due to encounters with single stars around the 10\%
   lagrangian radius. Escapers with high energies come from encounters
   with binaries in the cluster cores.}
\label{esc1}
\end{figure}

In our simulations, we checked the energy of each star when its regular force in NBODY6 was calculated.
In order to do this,
potential energies were calculated with respect to all stars and the kinetic
energy was calculated relative to the cluster velocity from the time of the last data output.
The time and radius where stars acquired positive energies were noted and we will refer to this
radius as the 'scattering distance' $R_{Scatt}$.
If stars still had positive energies after their distance from the cluster center had increased by 
a factor of two compared to the scattering distance, we assumed that the star was successfully scattered
out of the cluster and measured its excess energy above the $E = 0$ energy threshold. Stars that fell 
below $E = 0$ before they reached twice the scattering distance were not counted as escapers.
\begin{figure}
\epsfxsize=8.3cm
\begin{center}
\epsffile{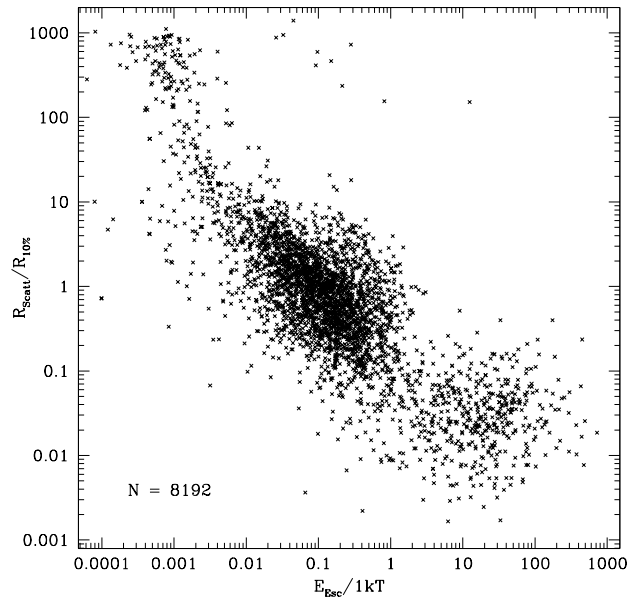}
\end{center}
\caption{Same as Fig.\ \ref{esc1} for the $N = 8192$ cluster. The distribution is similar to the
$N = 512$ case, except that escapers from binary encounters are coming from much smaller distances.}
\label{esc2}
\end{figure}

Fig.\ \ref{esc1} shows, as an example for a low-$N$ cluster, the distribution of energies against 
scattering radius for the $N = 512$ star models.
The energies are expressed in terms of the one-dimensional velocity dispersion of all bound stars
$1kT \; = \; 2/3<\!\!E_{Star}\!\!>$
and the distances are measured in units of the 10\% radius. One can notice three different
concentrations in this diagram. Stars escaping due to single-star encounters have typical energies
of a few tenths of 1kT and acquire the escape energies at distances around the 10\% lagrangian radius.
Single-star escapers coming from distances closer to the cluster center have higher energies
since stars at smaller radii move with larger velocities and can therefore also create higher
energy escapers. The most distant single star escapers come from around  $R = 10 \; R_{10\%}$,
which correspond roughly to the half-mass radius.
Stars escaping due to single-star encounters form the majority of escapers. 
\begin{figure*}
\epsfxsize=17cm
\begin{center}
\epsffile{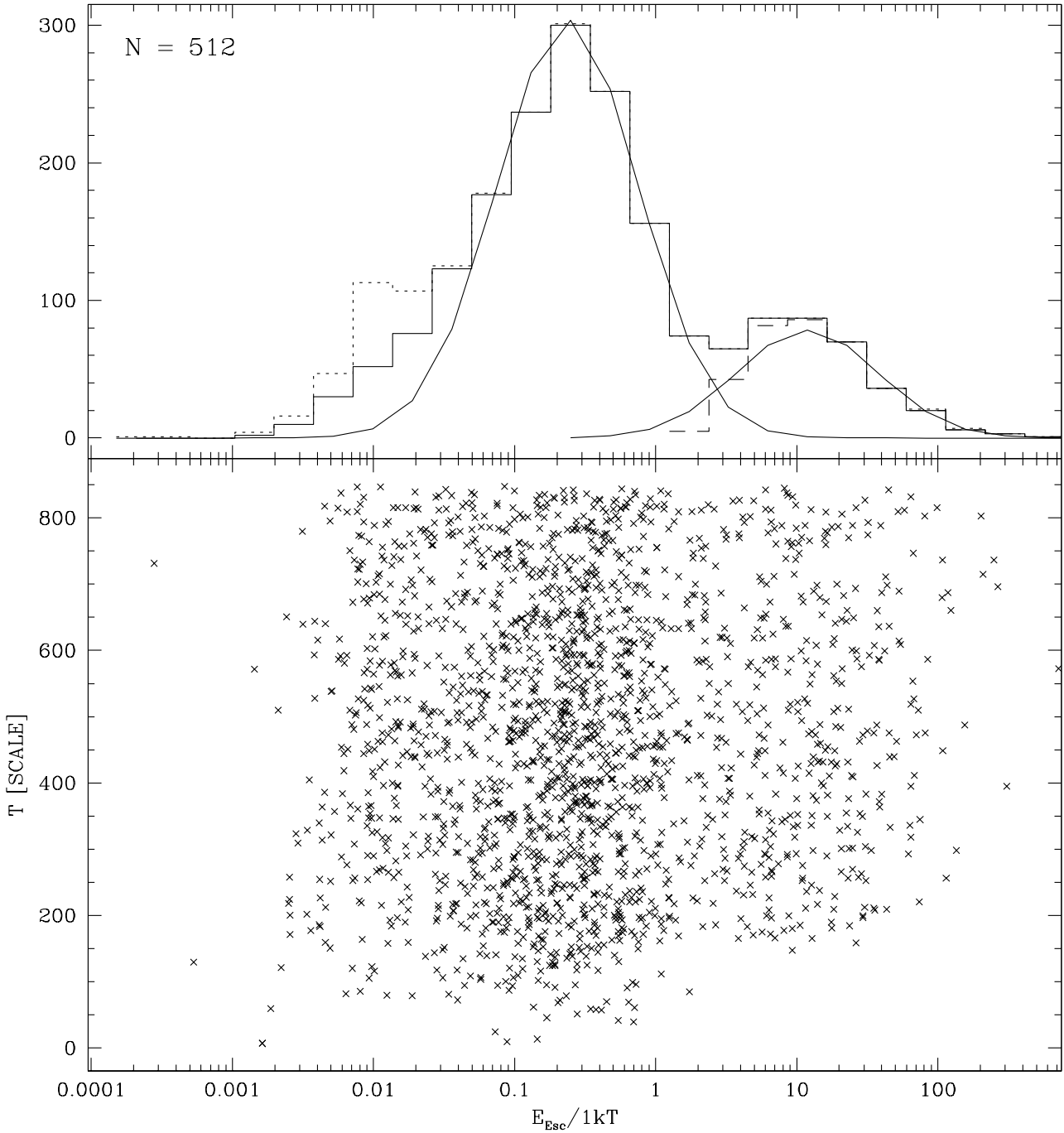}
\end{center}
\caption{Distribution of escape energies as a function of scaled time for clusters with $N = 512$
stars. Time is scaled according to eq.\ 1. The dotted
 line shows the distribution of all stars, the solid line shows the distribution after stars that
  escape due to the movement of the cluster have been subtracted. Escapers due to single
  star encounters follow roughly a gaussian distribution. The dashed line shows the remaining
 distribution after this gaussian has been subtracted from the high-energy escapers. Binary induced
  escape sets in only after core collapse and also follows roughly a gaussian distribution in energy.}
\label{esc3}
\end{figure*}

Stars which escape because of binary-single star interactions are created only in the cluster cores
and have typical energies between 2 and 100 kT. Since the energy of the escaper is coming from the
internal energy of the binary, there is no dependence of the escape energy on the 
scattering radius. Although smaller in number, escapers due to 
binary encounters carry away more than 95\% of the total energy of escapers.

A third group of stars is created in the halos at distances larger than 10 half-mass radii.
These stars escape since the cluster center moves erratically as a result of the recoil from high 
energy encounters in the core, which will unbind loosely bound stars.

A similar plot for the 8K model (Fig.\  \ref{esc2}) shows that the energy of the single star escapers
changes very little with particle number. In addition, the mean scattering distance remains the
same. Escapers due to binaries are now well separated from the single star
escapers. Their mean distance from the center has decreased by almost a factor of 10 due to the
more concentrated core of the high-$N$ model. 
\begin{figure*}
\epsfxsize=17cm
\begin{center}
\epsffile{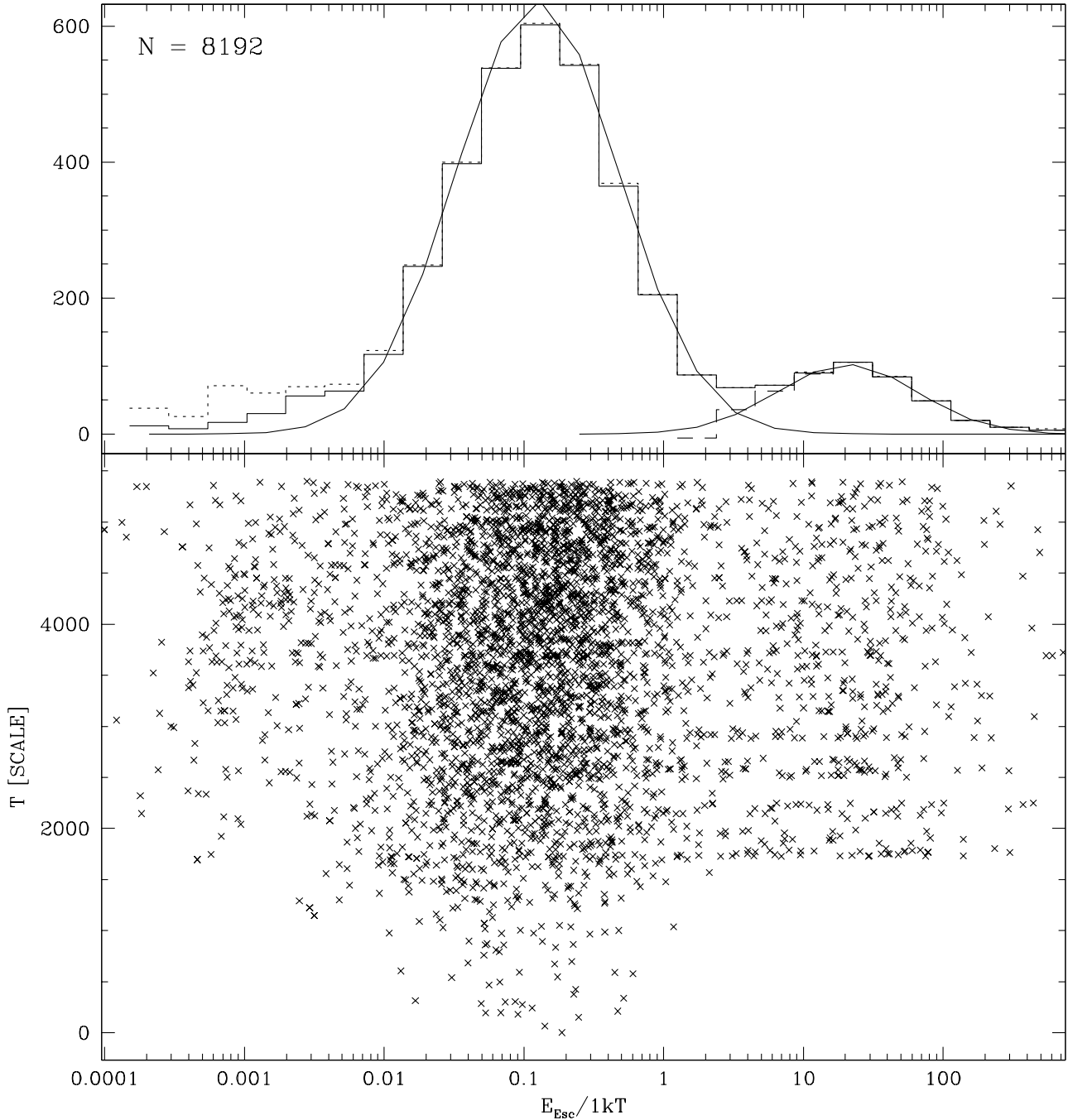}
\end{center}
\caption{Same as Fig.\ \ref{esc3} for $N = 8192$ stars. The distribution of escapers is very
 similar to the $N = 512$ case.  However, stars that escape due to encounters
  with binaries do not escape continuously, but in several distinct phases which are related to the
 core oscillations of the cluster.}
\label{esc4}
\end{figure*}

Fig. \ref{esc3} shows the distribution of escape energies as a function of scaled time 
(calculated according to eq.\ 1) for $N = 512$.
Escape of high-energy stars sets in only after core collapse (happening at around T = 150
$N$-body time units). In agreement with Giersz \& Heggie (1994a), we find that before core collapse,
stars escape due to single star encounters only. 
The typical escape energies of single-star escapers show no variation with time and lie between 0.04 and 1kT. 
After stars escaping due to the movement of the cluster are taken out, the energy 
distribution can be fitted by a sum of two gaussians, 
which correspond to the two escape mechanisms.
The majority of stars, typically about
80\% to 85\%, escape due to single-star encounters. 

Fig.\ \ref{esc4} shows the escape energies and escape times for stars in the 8K model. The energy
distribution of stars escaping due to single-star and binary encounters is similar to the 
$N = 512$ case. Escapers by binary encounters tend to have slightly higher energies than in
the $N=512$ star case. 
The fraction of binary escapers is slightly smaller than in the $N =512$ case
because the fraction of binaries compared to the number of cluster stars has decreased. 
The most striking difference is that in the 8K model, stars from binary encounters
do not have a smooth distribution in time, but escape in several distinct phases. 

\begin{figure}
\epsfxsize=8.3cm
\begin{center}
\epsffile{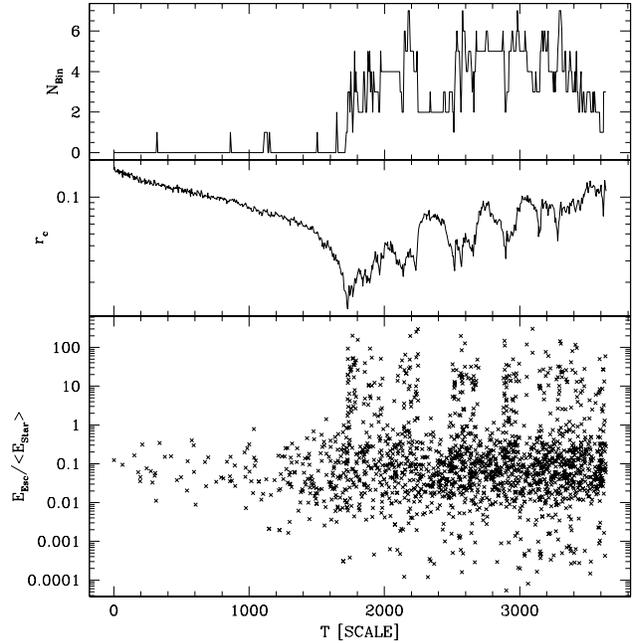}
\end{center}
\caption{Evolution of the escape energies (bottom panel), core radius (middle) and binary number (top) 
 in the 8K model. Although binaries are present all the time, they are efficient in creating escapers
  only when the cluster is in core collapse.}
\label{esc5}
\end{figure}

Fig. \ref{esc5} 
shows that these phases are linked to the core oscillations of the 8K model. Stars escape by 
binary encounters only when the cluster is in core collapse since the core density
is much lower in an expansion phase where close encounters between binaries and 
single stars are unlikely. Stars escaping because of
single-star encounters show much less variation with time as they are typically created at
larger radii that change less due to the oscillations. Fig.\ \ref{esc4} also shows that the rate
at which single-star escapers are created is increasing during and after core collapse which
could be the result of the increasing anisotropy. As a consequence, more loosely bound halo stars will 
travel through high-density regions where they have a high chance of being scattered to escape
energies. In addition, the overall change in the density distribution can also increase the
escape rate.  

\subsection{Binaries}   

\begin{figure}
\epsfxsize=8.3cm
\begin{center}
\epsffile{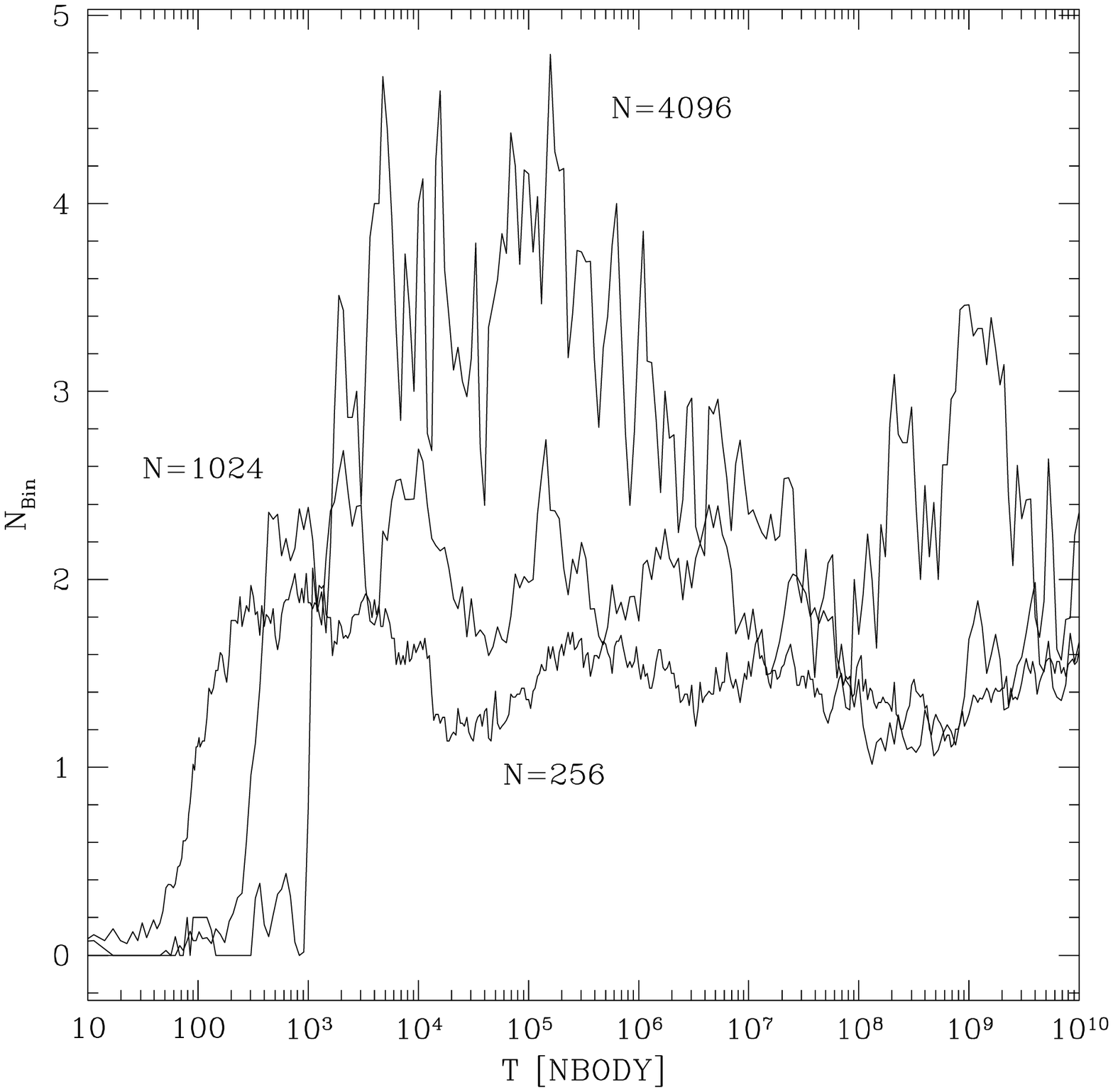}
\end{center}
\caption{Number of bound, regularised binaries as a function of time.
 The binary number reaches a maximum after about 10 core collapse times. It is larger for high-$N$ clusters and decreases
 as the clusters lose stars.}
\label{bin1}
\end{figure}

Fig.\ \ref{bin1} shows the number of bound binaries as a function of time. We 
assume that the number of binaries is equal to the number of regularised pairs in NBODY6, which
should be the case except for transient encounters between cluster stars which are also regularised
if they are close enough. 

No binaries are present at the start of our simulations and they are created only during and after 
core collapse. The maximum number of bound binaries is reached after about 10 core collapse times.
Throughout the post-collapse expansion, a few binaries are sufficient to drive the
expansion of the clusters, even for clusters with several thousand stars.

The number of binaries increases with the number of cluster stars, while their fraction decreases.
If the number of binaries is taken at the maximum value, the increase can be fitted approximately as 
$N_{Bin} \propto N^{0.3}$. This is comparable to the increase given by Giersz \& Heggie (1994b), who found 
$N_{Bin} \propto N^{0.2}$ after a few core collapse times had passed. Given the considerable scatter
in our data and the fact that the binary number in
our models is still rising after a few core collapse times, the agreement is satisfactory.   

Comparison of the members of binary systems at successive times shows that the vast majority of them
are not transient encounters but bound systems, especially for clusters with high particle numbers.
In addition, the number of binaries in the core is increasing with the particle number
in a similar way than the total binary number.
Hence the number of active binaries needed to support the expansion of the clusters 
seems to rise with $N$, in contrast to Goodman's (1984) results from gaseous model
calculations.

For later times, the binary fraction starts to drop again as the clusters contain fewer and fewer 
stars. This effect can best be seen for the high-$N$ runs.  
Runs with small particle numbers show almost no decrease, probably because at least one binary has to present in the cluster
core to drive the evolution. If this star is ejected, a new binary will be formed quickly, so a 
value of one forms a lower limit for the number of binaries (cf. Fig.\ 15). 

\begin{figure}
\epsfxsize=8.3cm
\begin{center}
\epsffile{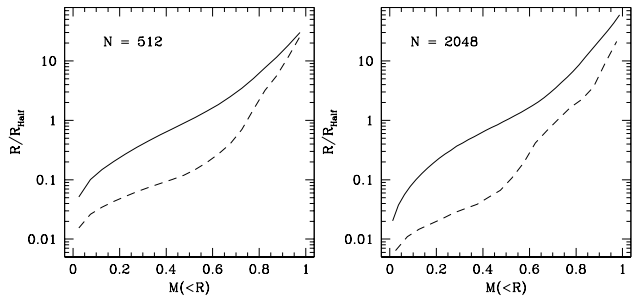}
\end{center}
\caption{Radial distribution of binaries and single-stars for clusters containing $N = 512$ and $N = 2048$ stars
 respectively. 
 Binaries (dashed) are more concentrated than single stars (solid line). In addition, binaries in high-$N$ clusters
  can be found at smaller radii.} 
\label{bin2}
\end{figure}

The radial distribution of bound binaries is shown in Fig.\ \ref{bin2}. Although more concentrated then single-stars,
a significant fraction of binaries can be found outside the core. Only one quarter of all binaries is 
located within the 3\% lagrangian radius. Binaries outside the core are mostly ejected 
by three and four body interactions and move on very elongated orbits. Subsequent passages through
the cluster center will remove them from the cluster. Given the smaller density
of stars and binaries in the halo, binary activity should be confined to the cluster centers.

\subsection{Anisotropy}

\begin{figure}
\epsfxsize=8.3cm
\begin{center}
\epsffile{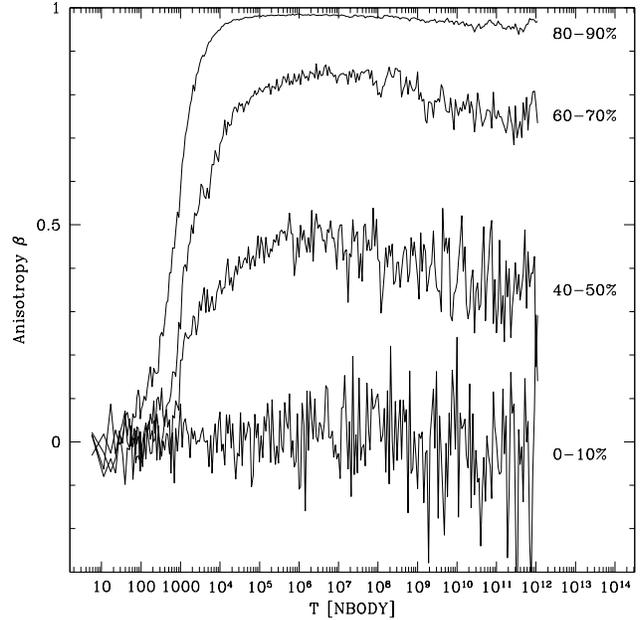}
\end{center}
\caption{Evolution of the anisotropy with time for 4 different radii for clusters with $N=4096$ stars.
   Clusters are strongly anisotropic in their outer parts but remain isotropic near their
    centers. The mean anisotropy increases up to $10^6$ $N$-body time units and decreases slowly 
     afterwards. A similar behavior is found for other particle numbers.}    
\label{an1}
\end{figure}

Fig. \ref{an1} shows the evolution of the mean anisotropy for clusters with $N = 4096$ stars.  
To calculate the anisotropy, only bound stars are used. 
Following Binney \& Tremaine (1987), we define the anisotropy parameter $\beta$ as 
$\beta = 1 - (\sigma_t^2/2 \sigma_r^2)$, which is half as large as the $A$-parameter
used by Giersz \& Spurzem (1994) and Takahashi (1996). The anisotropy rises from the start
of the simulations and reaches a maximum after about $10^6$ $N$-body time units, corresponding
to $1.5 \cdot 10^4$ initial relaxation times. We find that the same number of initial relaxation
times is required for other $N$. We also find a close agreement between the time when the 
clusters reach maximum anisotropy and the time when their halos reach maximum expansion 
relative to the inner radii. At this time, the outermost shells have become nearly 100\% 
radially anisotropic. Clusters remain isotropic inside their 20\% radius throughout the simulation,
which agrees very well with results found by Takahashi (1996) from Fokker-Planck
simulations. 

One can see that the anisotropy is decreasing for very late times, which is
the result of the increasing removal of weakly bound stars as the particle numbers drop.
Stars with energies near $E=0$ are predominantly on near radial orbits when closer
to the cluster center. The removal of such stars must therefore decrease the anisotropy
at intermediate radii.

\begin{figure}
\epsfxsize=8.0cm
\begin{center}
\epsffile{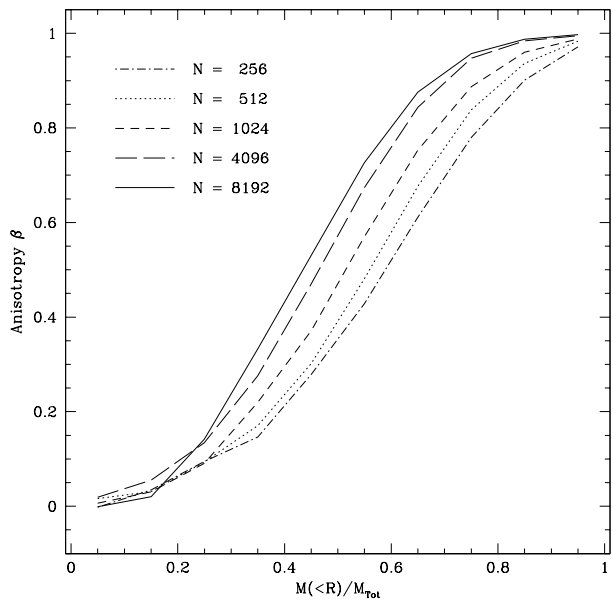}
\end{center}
\caption{Maximum anisotropy in dependence of the initial number of cluster stars.
 The anisotropy is plotted as a function of enclosed mass. Shown are curves from 
  $N = 256$ (bottom) to $N=8192$ (top). The anisotropy in the halo increases with 
  increasing particle number.}     
\label{an2}
\end{figure}

We can compare the anisotropy profile for clusters with different particle numbers by binning data 
from several snapshots
around the time when the clusters reach maximum anisotropy.  Fig. \ref{an2} shows the resulting
profiles as a function of enclosed mass. The anisotropy rises monotonically from the core, which is 
isotropic for all $N$, towards the halo. For large $N$, the differences become smaller and it might
be possible that the profiles are converging towards a maximum profile at the high-$N$ end.
Unfortunately it is difficult to say if this is really the case, due to the small number of
simulations made (only one for $N=8192$) and the resulting uncertainties in the profiles. If so, 
it would mean that there is an universal density and velocity profile, which only clusters with
large enough $N$ can reach, since the evolution of low-$N$ models towards this profile is 
interrupted by the removal of their outer halo.

\begin{figure}
\epsfxsize=8.3cm
\begin{center}
\epsffile{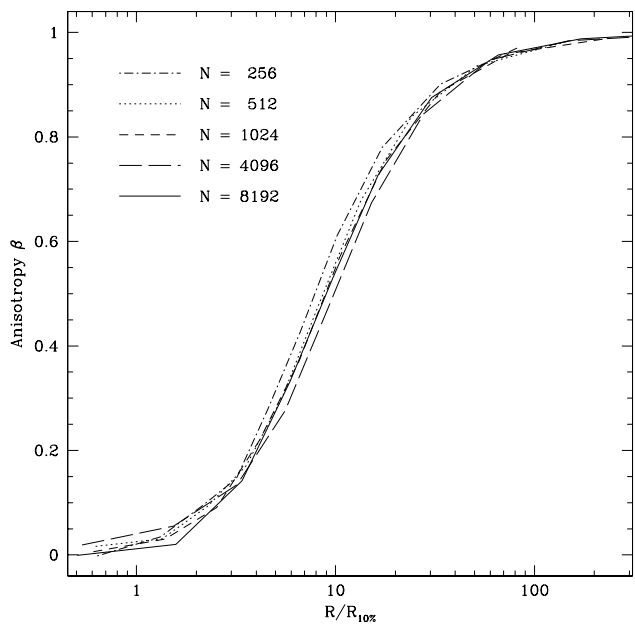}
\end{center}
\caption{Maximum anisotropy as a function of distance to the cluster center. If distances are
 scaled by the 10\% lagrangian radius, the anisotropy profile is independent of the particle number.
 Most stars should therefore be scattered into the halo from around this radius.} 
\label{an3}
\end{figure}

It was already shown that the cluster expansion is driven by encounters of stars near the 10\% 
lagrangian radius (Fig. \ref{rrel}). Hence, the 10\% lagrangian radius should also play an
important role for the anisotropy profile. Fig. \ref{an3} depicts the anisotropy profile for 
different particle numbers, calculated by dividing the radii of each model by its 10\% 
lagrangian radius. It can be seen that 
the profiles are nearly the same for all $N$, showing again that the 10\% radius
is the radius where halo stars are created. We found that no other radius leads to such an
agreement in the profiles.        

H\'enon (1973) found that
spherical stellar systems with radially anisotropic distribution functions can be vulnerable to 
radial orbit instabilities. His results were later confirmed and extended by Barnes, Hut
\& Goodman (1986) and Dejonghe \& Merritt (1988). Dejonghe \& Merritt (1988) found that models
based on Plummer's density law become unstable if the global anisotropy exceeds a 
critical value of $2 T_r/T_t \approx 2.0$. The anisotropies in our highest-$N$ models are close 
to this value, and so it might be interesting to look for signs of such instabilities in our 
runs.

We calculated for all models with $N \ge 1024$ and all times when data was stored the moment of 
inertia tensor of the stellar distribution and its eigenvectors and eigenvalues. We studied 
only radial shells between the 20\% and 80\% lagrangian radius, since the velocity distribution is
isotropic for smaller radii. Since previous work indicates that radially anisotropic stellar systems 
develop bar-like mass distributions, we first checked for stellar bars in our models.

\begin{figure}     
\epsfxsize=8.3cm
\begin{center}
\epsffile{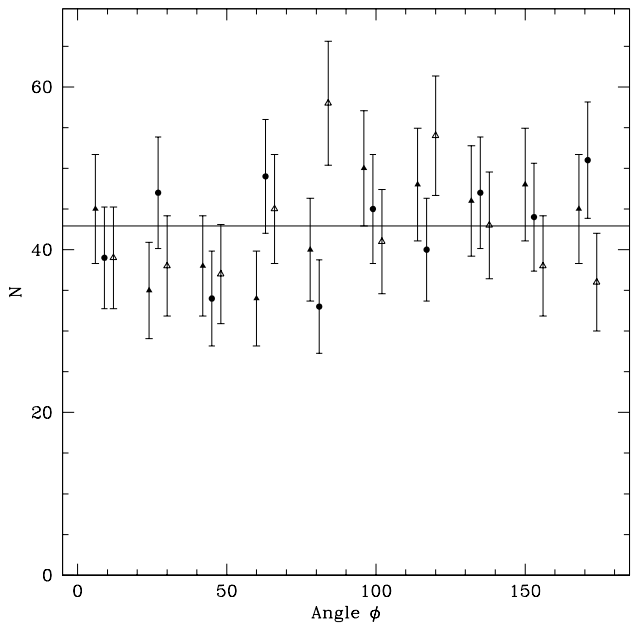}
\end{center}
\caption{Angular distribution of the eigenvectors that belong to the smallest eigenvalue
  for the 8K run. Circles
 show the projection into the x-y plane, open triangles into the x-z plane and filled triangles into
  the y-z plane. The solid line shows the expected value for a uniform distribution. 
  The distribution of the data from the $N$-body run is compatible with a random distribution.} 
\label{oi2}
\end{figure}

Fig.\ \ref{oi2} shows for the 8K run the angular distribution  
of the eigenvectors which belong to the smallest moment of inertia. 
It can be seen that their distribution is more or less random.
This is not what one would expect in the 
presence of radial orbit instabilities, since bar-like configurations should have fixed 
axes in space, at least for a certain amount of time. In a second test we compared the ratio of the smallest
eigenvalue to the sum of the other two. 
If a cluster collapses into a bar-like configuration, one would expect to see a sudden drop
of this ratio. However, no such drop could be seen anywhere in the run.
Hence no signs of instabilities 
could be found in the 8K run and similar results were obtained for runs with smaller particle numbers.
We also found no signs for disc-like instabilities in any of our runs.

There are several possible explanations for the absence of radial orbit instabilities:
Our distributions might not be anisotropic enough to become unstable, or relaxation processes
suppress them since they tend to make distributions spherical. Most important for halo stars 
are passages through the inner cluster parts, where encounters with other stars can significantly 
change their orbits.
As was shown earlier (Fig.\ 3), the relaxation times in the center change only slowly 
with the particle number since 
high-$N$ clusters are more concentrated. The onset of radial orbit instabilities might
therefore be suppressed by relaxation processes in the cluster center, no matter how large
$N$ becomes.

A third possibility is that the halo is constantly renewed and extended by stars ejected from the 
cluster center. Since such stars will be scattered uniformly in all directions,
they can also prevent the formation of structures in the halo.  

\section{Conclusions}

We have performed a set of $N$-body simulations of isolated single-mass clusters starting from Plummer 
profiles. Our main focus was on the post-collapse evolution, and for   
the first time we could follow this evolution until nearly complete disintegration of the clusters,
thereby extending the parameter space of collisional stellar dynamics significantly.

We found that after a sufficient number of relaxation times has passed, the structure of the clusters 
becomes independent
of the initial density profile and particle number, and can be characterized
by just two parameters, as e.g.\ the actual particle number and the half-mass radius.
For very large $N$, the cluster structure might become independent of the actual particle number too,
although our simulations could not reach large enough $N$ to test this assumption.
We found that isolated clusters evolve along a single sequence of models and runs starting from different
particle numbers can be stitched together. 
This can be relevant for the post-collapse evolution of galactic globular 
clusters which have half-mass radii much smaller than their tidal radii, since these systems would be 
nearly isolated and might have profiles similar to our models. 
 
We found that low-$N$ clusters have larger radii after core collapse, so that all clusters have similar
relaxation times during the post-collapse expansion phase.   
As a consequence, the mass-loss rates depend only weakly on the particle number.
Due to their higher concentration, high-$N$ clusters loose mass even faster than
low-$N$ models.

Stars in isolated clusters are ejected almost entirely from within the half-mass radius, with encounters between
single stars being the dominant escape mechanism. We could establish a clear correlation between the 
excess energy of escapers and the position where the escapers are created. Such a relation could
also exist for clusters in tidal fields, although the long time necessary for escape complicates
the escape of stars in this case (Fukushige \& Heggie 2000, Baumgardt 2001).

For all clusters, only a few binaries are present and drive the expansion. Their fraction decreases
with increasing particle number, while the total number increases roughly as $N_{Bin} \sim N^{0.3}$.
About 15\% of the stars are ejected due to encounters with binaries and these carry away a large fraction
of the cluster energy. 
Binaries are strongly concentrated towards the cluster cores, and the ejection by binaries also happens only  
near the cluster centers.
For clusters showing core oscillations, binary induced escape is efficient only in the contraction phases of the
core, while the escape due to single-star encounters fluctuates much less with the 
core oscillations.
 
A radially anisotropic velocity distribution is created in the cluster halos, mainly as a result of two-body
encounters near the 10\% lagrangian radius. High-$N$ clusters have smaller core sizes and show therefore
larger overall anisotropies. It seems conceivable that isolated clusters are vulnerable to radial orbit 
instabilities for sufficiently large $N$, but no indication for such
instabilities could be found.
 
\section*{Acknowledgments}
We are grateful to Rainer Spurzem for his help with the NBODY6++ code and to Jun Makino for useful
discussions. We would also like to thank an anonymous referee for comments which improved the presentation
of the paper.  
It is a pleasure to acknowledge the support of 
the European Commission through TMR grant number ERB FMGE CT950051 (the TRACS  
Programme at EPCC). The parallel computations were performed on the
CRAY T3E of HLRS Stuttgart.
H.B.\ was supported by PPARC under grant 1998/00044.

\bsp
\label{lastpage}


\begin{thebibliography}{00}
\bibitem[1971]{a71} Aarseth S., 1971, Ap\&SS 13, 324 
\bibitem[1999]{a99} Aarseth S., 1999, PASP 111, 1333   
\bibitem[1962]{a62} Antonov V.\ A., 1962, Vestnik Leningrad Univ., 7, 135
\bibitem[1986]{bhg} Barnes J., Hut P., Goodman J., 1986, ApJ 300, 112 
\bibitem[2001]{me} Baumgardt H., 2001, MNRAS 325, 1323    
\bibitem[1987]{bt} Binney J., Tremaine S., 1987, Galactic Dynamics, Princeton Univ. Press, Princeton
\bibitem[1994]{br} Breeden J.\ L., Cohn H.\ N., Hut P., 1994, ApJ 421, 195
\bibitem[1999]{ch85} Casertano S., Hut P., 1985, ApJ 298, 80
\bibitem[1985]{c85} Cohn H., 1985, in Goodman J., Hut P. (eds), IAU Symposium 113,
 Dynamics of Star Clusters, Kluwer, Dordrecht, p.\ 161  
\bibitem[1988]{dm} Dejonghe H., Merritt D., 1988, ApJ 328, 93      
\bibitem[1999]{dru} Drukier G.\ A., Cohn H.\ N., Lugger P.\ M., Yong H., 1999, ApJ 518, 233  
\bibitem[2000]{fh} Fukushige T., Heggie D.C., 2000, MNRAS 318, 753   
\bibitem[1994]{gha} Giersz M., Heggie D.\ C., 1994a, MNRAS 268, 257     
\bibitem[1994]{ghb} Giersz M., Heggie D.\ C., 1994b, MNRAS 270, 298
\bibitem[1997]{gh7} Giersz M., Heggie D.\ C., 1997, MNRAS 286, 709
\bibitem[1994]{gs} Giersz M., Spurzem S., 1994, MNRAS 269, 241 
\bibitem[2000]{gs2}  Giersz M., Spurzem S., 2000, MNRAS 317, 581 
\bibitem[1984]{jg} Goodman J., 1984, ApJ 280, 298
\bibitem[1987]{jg87} Goodman J., 1987, ApJ 313, 576
\bibitem[1975]{hh} Heggie D.\ C., 1975, MNRAS 173, 729
\bibitem[1960]{he60} H\'enon M., 1960, Ann.\ Astrophys.\ 23, 668    
\bibitem[1973]{h73} H\'enon M., 1973, A\&A 24, 229
\bibitem[1983]{il} Inagaki S., Lynden-Bell D., 1983, MNRAS 205, 913     
\bibitem[1970]{l70} Larson R.\ B., 1970, MNRAS 150, 93 
\bibitem[1978]{ls} Lightman A.\ P., Shapiro S.\ L., 1978,  Rev.\ Mod.\ Phys.\ 50, 437
\bibitem[1980]{le} Lynden-Bell D., Eggleton P.\ P., 1980, MNRAS 191, 483 
\bibitem[1968]{lw} Lynden-Bell D., Wood R., 1968, MNRAS 138, 495
\bibitem[2001a]{mj1} Makino, J., 2001, in {\it Dynamics of Star Clusters and the Milky Way},
   eds.\ S.\ Deiters, B.\ Fuchs, A.\ Just, R.\ Spurzen, R.\ Wielen, ASP Conference Series 228, p.\ 87 
\bibitem[2001b]{mj2} Makino J., 2002, in {\it Astrophysical Supercomputing using Particle Simulations},
  eds.\ J.\ Makino, P.\ Hut, IAU Symposium 208, in preparation
\bibitem[1992]{ma} Makino J., Aarseth S.J., 1992, PASJ 44, 141   
\bibitem[1972]{s72} Spitzer L.\ Jr., Shapiro S.\ L., 1972, ApJ 173, 529
\bibitem[1987]{s87} Spitzer L.\ Jr., 1987, Dynamical Evolution of Globular Clusters,
  Princeton University Press, Princeton
\bibitem[1999]{s99} Spurzem R., 1999, in Riffert H., Werner K. (eds),
 Computational Astrophysics, The Journal of Computational and Applied Mathematics
   (JCAM) 109, Elsevier Press, Amsterdam, p.\ 407
\bibitem[1996]{sa} Spurzem R., Aarseth S.\ J., 1996, MNRAS 282, 19
\bibitem[2002]{sb} Spurzem R., Baumgardt H., 2002, MNRAS submitted
\bibitem[1995]{t95} Takahashi K., 1995, PASJ 47, 561
\bibitem[1996]{t96} Takahashi K., 1996, PASJ 48, 691
\bibitem[2000]{t00} Takahashi K., Portegies Zwart S.\ F., 2000, ApJ 535, 759
\end{thebibliography}
\end{document}